\documentstyle[12pt]{article}
\newlength{\pubnumber} 

\catcode`\@=11

\def\section{\@startsection{section}{1}{\z@}{3.5ex plus 1ex minus .2ex}
 {2.3ex plus .2ex}{\large\bf}}
\def\subsection{\@startsection{subsection}{2}{\z@}{2.3ex plus .2ex}
 {2.3ex plus .2ex}{\bf}}

\begin{document}

\begin{titlepage}
\samepage{
\setcounter{page}{1}
\rightline{RU-96-33}
\rightline{IASSNS-HEP-96/48}
\rightline{\tt hep-th/9607190}
\rightline{July 1996}
\vfill
\begin{center}
 {\Large \bf On the Worldsheet Formulation of
       the Six-Dimensional Self-Dual String\\}
\vfill
 {\large Philip C. Argyres$^1$\footnote{E-mail address:
       argyres@physics.rutgers.edu}
       $\,$and$\,$ Keith R. Dienes$^2$\footnote{E-mail address:
       dienes@sns.ias.edu}\\}
\vspace{.19in}
 {\it $^1$ Department of Physics and Astronomy, Rutgers University\\
       Piscataway, New Jersey 08855~~\\}
\vspace{.03in}
 {\it $^2$ School of Natural Sciences, Institute for Advanced Study\\
       Olden Lane, Princeton, New Jersey 08540~~\\}
\end{center}
\vfill
\begin{abstract}
  {\rm  Despite recent evidence indicating the existence of
      a new kind of self-dual six-dimensional superstring,
      no satisfactory worldsheet formulation of such
      a string has been proposed.
      In this note we point out that
      a theory built from ${\bf Z}_4$ parafermions may have the
      right properties to describe the light-cone conformal
      field theory of this string.
      This indicates a possible worldsheet formulation of this theory.}
\end{abstract}
\vfill}
\end{titlepage}

\setcounter{footnote}{0}
\def\beq{\begin{equation}}
\def\eeq{\end{equation}}
\def\beqn{\begin{eqnarray}}
\def\eeqn{\end{eqnarray}}
\def\calZ{{\cal Z}}
\def\half{{\textstyle {1\over2}}}
\def\bone{{\bf 1}}
\def\ie{{\it i.e.}}
\def\eg{{\it e.g.}}
\def\thetatwo{{\vartheta_2}}
\def\thetathree{{\vartheta_3}}
\def\thetafour{{\vartheta_4}}
 \font\cmss=cmss10 \font\cmsss=cmss10 at 7pt
\def\IZ{\relax\ifmmode\mathchoice
 {\hbox{\cmss Z\kern-.4em Z}}{\hbox{\cmss Z\kern-.4em Z}}
 {\lower.9pt\hbox{\cmsss Z\kern-.4em Z}}
 {\lower1.2pt\hbox{\cmsss Z\kern-.4em Z}}\else{\cmss Z\kern-.4em Z}\fi}

\hyphenation{pa-ra-fer-mion pa-ra-fer-mion-ic pa-ra-fer-mions }
\hyphenation{su-per-string frac-tion-ally su-per-re-pa-ra-met-ri-za-tion}
\hyphenation{su-per-sym-met-ric frac-tion-ally-su-per-sym-met-ric}
\hyphenation{space-time-super-sym-met-ric fer-mi-on}
\hyphenation{Ne-veu-Schwarz-like Ramond-like}


\setcounter{footnote}{0}
\section{Introduction}

Over the past year, evidence \cite{Wit,Strom,Wit2,Ganor1}
has accumulated for the existence of a new kind of closed superstring which
\begin{itemize}
\item  has six-dimensional Lorentz invariance;
\item  is $N=(2,0)$ spacetime supersymmetric;  and
\item  whose massless modes are in a tensor supermultiplet.
\end{itemize}
This last property implies that this string has no graviton
in its spectrum, and that it couples to an antisymmetric two-form
$B_{\mu\nu}$ with self-dual field strength.
This string, which is the first example of non-trivial
infrared dynamics of a quantum theory in six flat dimensions,
arises in certain compactifications of Type~IIB strings or of $M$-theory
when parallel five-branes coincide.
There are also self-dual strings with $N=(1,0)$ supersymmetry in six
dimensions \cite{DMW,GH,SeiWit,Ganor2}, as well as $N=2$
supersymmetric strings in four dimensions \cite{Wit3,Kleb}.

The above properties, which concern only the macroscopic behavior of the
self-dual string, can be summarized by noting \cite{DVV} that
the massless states of the self-dual string are the same as those of
the classical six-dimensional Green-Schwarz superstring.  (Recall that
the six-dimensional Green-Schwarz string is indeed
consistent at the classical level.)
A microscopic description of the self-dual string, on the other hand,
is lacking.

In this note we will {\it assume}\/ that there exists a microscopic
worldsheet formulation of the self-dual string, and seek to construct it.
We should emphasize at the outset, however, that there is very little
evidence to support the existence of such a microscopic formulation.
Indeed, given that the self-dual string is necessarily strongly
coupled (by virtue of its self-duality), it is hard to guess
the properties of a microscopic description.
In Ref.~\cite{Schwarz}, for example, it is argued
that a worldsheet formulation of the strongly-coupled string
might exist because microscopic string states in six flat dimensions
do not couple to the $B_{\mu\nu}$ field at zero momentum.  However,
such states are derivatively coupled, which would seem to
imply only that the effective six-dimensional theory of
the massless tensor multiplet is weakly coupled.

One clue to such a microscopic description comes from studies of black
hole entropy, from which it is argued \cite{DVV} that the self-dual
string
\begin{itemize}
\item  has an asymptotic degeneracy of BPS-saturated
     states consistent with a light-cone gauge worldsheet theory with
     central charge $c=6$.
\end{itemize}
Indeed, in Ref.~\cite{DVV}, the use
of a weakly-coupled description of this string
was justified by noting that the
counting of BPS states should not change as the theory is
deformed to weak coupling.
This then gives rise to a worldsheet formulation of the self-dual string,
which the authors of Ref.~\cite{DVV} took
to be that of the six-dimensional Green-Schwarz string.   This theory
is described by
a light-cone gauge conformal field theory (CFT) consisting of four
transverse bosons $X^i$ and four worldsheet fermions $S^a$ carrying
spacetime spinor indices in the $2(\bf 2,1)$ representation of the
$SU(2)\times SU(2)$ massless little group.  It is well-known, however, that
this theory is not Lorentz invariant.  Thus, perhaps, this worldsheet
formulation describes the self-dual string in some background
in which six-dimensional Lorentz invariance is broken but in which the string
is weakly coupled.

Indeed, the mere existence of the
classical Green-Schwarz actions in six, four, and three dimensions
might be taken as an additional piece of evidence of a microscopic
worldsheet description of the self-dual string.
In particular, the Green-Schwarz actions are ``special''
in the sense that they depend on the existence of a non-trivial
$\kappa$-symmetry.

The purpose of this note is to give evidence for another,
closely related worldsheet description of the self-dual string.
The main guideline we follow is that the central charge
of this worldsheet theory in light-cone gauge should be $c=6$.
Our procedure will be as follows.
First, we will construct
a partition function which has all of the properties necessary
for a consistent interpretation as that of the self-dual string.
Then, we will show how the
formulation of this partition function
naturally suggests a construction of the associated worldsheet
conformal field theory.

\section{Constructing the Partition Function}

Let us begin by reviewing the partition function of the
six-dimensional Green-Schwarz string.
This partition function is given by
\beq
     Z(\tau)~=~  ({\rm Im}\, \tau)^{-2}\, \biggl |
        f_B(\tau) - f_F(\tau) \biggr |^2~
\label{expectedpartfunct}
\eeq
where the bosonic and fermionic contributions $f_B(\tau)$ and
$f_F(\tau)$ are equal and given by:
\beq
    f_B(\tau)~=~f_F(\tau)~=~
          4\, \prod_{n=1}^\infty\, \left({{1+q^n}\over{1-q^n}}\right)^{4}~,
	\qquad q\equiv e^{2\pi i \tau}~.
\label{infprod}
\eeq
This infinite product can be written in terms
of Jacobi $\vartheta_i$-functions and the Dedekind $\eta$-function as
\beq
      4\,\prod_{n=1}^\infty \, \left({{1+q^n}\over{1-q^n}}\right)^{4} ~=~
        {\thetatwo^2(\tau) \over \eta^6(\tau)} ~.
\label{prodtheta}
\eeq
These functions are
the characters of four worldsheet bosons and four Majorana-Weyl worldsheet
fermions, which comprise the
$c=6$ light-cone gauge field content of the six-dimensional Green-Schwarz
string.

The worldsheet formulation described in Ref.~\cite{DVV}
then proceeds from this point, assuming such bosonic and fermionic
fields on the worldsheet.  However,
as further analyzed in Ref.~\cite{Schwarz},
this approach leads to the well-known difficulties concerning
Lorentz invariance.  Indeed, the attempted solution to this
problem proposed in Ref.~\cite{Schwarz} involves increasing
the central charge beyond $c=6$.
By contrast, in this note we shall explore whether there exists an
alternative formulation of the theory which does not
involve simple worldsheet bosons and fermions,
but which nevertheless retains the properties we desire.

In general, the problem we face is that we wish
to realize or interpret the infinite product (\ref{infprod})
in terms of the underlying characters of an unknown six-dimensional theory
via an expression of the form
$f=\eta^{-4} \sum \chi_1 \chi_2 \chi_3 \chi_4$.  Here
$\eta^{-4}$ is the character of
four (transverse) free bosons,
the summation is over the different sectors of the hypothetical theory,
and $\chi_i$ are the characters of the worldsheet fields of that theory.
Six-dimensional Lorentz invariance suggests that each term in the expression
$f$ should have precisely four such $\chi_i$ characters in a light-cone
gauge formulation.
Moreover, spacetime supersymmetry requires that
there should exist two distinct ways of expressing the
infinite product (\ref{infprod})
in terms of the characters of that theory:  one combination of characters
$f_B$ should correspond to the spacetime bosons,
while the other set $f_F$  should correspond to the spacetime fermions.
For example, in the ten-dimensional Type~II string,
such inequivalent ways of writing the corresponding
infinite product would be $f_B=\thetathree^4-\thetafour^4$ and
$f_F=\thetatwo^4$;
these have equivalent product representations as a result of the usual
`abstruse' Jacobi identity.
By contrast, note that the {\it six}\/-dimensional Green-Schwarz string
would require terms of the form $f_F=\thetatwo^2$
and $f_B=(\thetathree^4-\thetafour^4)^{1/2}$;  the appearance
of the undesired square root in this expression ultimately signals
the breakdown of Lorentz invariance.

There exists an alternative realization of the infinite product
(\ref{infprod}) which has all of the desired properties \cite{AT,ADT}.
However, rather than relying on the characters of simple fermions,
this realization makes use of the characters of the $\IZ_4$ {\it parafermion}\/
conformal field theory \cite{ZKCFT}.
This $c=1$ conformal field theory,
which is equivalent \cite{Yang} to that of a free boson on a circle of radius
$\sqrt{3/2}$,
can be realized as the coset $SU(2)_4/U(1)$,
and has seven primary fields
denoted $\lbrace \phi^0_0, \phi^1_0, \phi^1_{\pm 1}, \phi^2_0, \phi^2_{\pm 1}
\rbrace$.
Here each primary field $\phi^j_m$ has been labelled by
its $SU(2)$ quantum numbers $j$ and $m$.
These primary fields are the parafermionic analogues of
the different toroidal boundary conditions of an ordinary Majorana-Weyl
fermion,
and they generalize the identity, fermion, and spin-fields of the Ising model.
The conformal dimension of each parafermionic primary field $\phi^j_m$
is given by $h^j_m=j(j+1)/6 - m^2/4$, and the corresponding character
will be denoted $\chi^{2j}_{2m}(\tau)$.
These characters,
which satisfy the identity $\chi^{2j}_{2m}=\chi^{2j}_{-2m}$,
are given by $\chi^{2j}_{2m}\equiv \eta c^{2j}_{2m}$
where $c^{2j}_{2m}(\tau)$ are the so-called
parafermionic {\it string functions} \cite{stringfunctions}.
As required, there exist two distinct linear combinations
$A^{\rm boson}$ and $A^{\rm fermion}$ of quadrilinear
products of these characters $\chi^{2j}_{2m}$ which
reproduce the desired infinite product \cite{ADT}:
\beq
      4\,\prod_{n=1}^\infty \, \left({{1+q^n}\over{1-q^n}}\right)^{4} ~=~
        A^{\rm boson}~=~
        A^{\rm fermion}~,
\label{prodA}
\eeq
where
\beqn
    A^{\rm boson}&=& \eta^{-4}\,\left\lbrack
       4(\chi^2_2)^4 ~-~32(\chi^2_2)(\chi^4_2)^3  \right\rbrack\nonumber\\
    A^{\rm fermion} &= & \eta^{-4}\,\left\lbrack
      4(\chi^0_0+\chi^4_0)^3(\chi^2_0)~-~4(\chi^2_0)^4   \right\rbrack ~.
\label{Abf}
\eeqn
Thus we are naturally led to a potential worldsheet formulation of the
self-dual string for which the worldsheet theory
consists not just of four transverse coordinate bosons, but also four $\IZ_4$
parafermions.  Each term in the expressions (\ref{Abf}) would then
correspond to a different bosonic or fermionic sector of this
underlying theory.

Since each $\IZ_4$ parafermion conformal field theory has central
charge $c=1$, this proposal would seem to imply a total worldsheet
light-cone central charge $c=8$, which disagrees with the desired value
$c=6$.  However, as guaranteed from the relations (\ref{prodA}), this
is ultimately not the case, for the minus signs within (\ref{Abf})
imply a cancellation of states in which the original $c=8$ Fock
space is ``deformed'' down to that of a $c=6$ theory \cite{ADT,AD,D}.
This delicate
cancellation is called an {\it internal projection}, and will be
discussed below.  Thus, we see that we have constructed a non-trivial
realization of a $c=6$ worldsheet theory in terms of a larger
parafermionic $c=8$ worldsheet theory, followed by an internal
projection.

This is not all, however.  In order to produce a fully consistent
theory, the total partition function must be modular invariant.
However, the naive partition function constructed from just the
expressions in (\ref{Abf}) is not modular invariant.  Rather, for
consistency, we find that we must add the additional contributions
\beqn
    B^{\rm boson} &= &  \eta^{-4}\,\left\lbrack
    8(\chi^0_0+\chi^4_0)^2(\chi^2_2)(\chi^4_2)
	      ~-~4(\chi^2_0)^2(\chi^2_2)^2 \right\rbrack \nonumber\\
    B^{\rm fermion} &=& \eta^{-4}\,\left\lbrack
    4(\chi^2_0)^2(\chi^2_2)^2 ~-~
	  16(\chi^0_0+\chi^4_0)(\chi^2_0)(\chi^4_2)^2 \right\rbrack ~,
\label{Bbf}
\eeqn
so that the total modular-invariant partition
function for this six-dimensional theory becomes
\beq
   Z(\tau) ~=~ ({\rm Im}\,\tau)^{-2}~
      \biggl\lbrace |A^{\rm boson}-A^{\rm fermion}|^2 ~+~
       3\, |B^{\rm boson}-B^{\rm fermion}|^2\biggr\rbrace~.
\label{partfunct} \eeq

On the face of it, the forced introduction of these extra $B$-sectors
has the potential to destroy the two features of our partition
function that we most wished to preserve:  spacetime supersymmetry,
and the central-charge reduction from $c=8$
to $c=6$.  Remarkably, however, the $B$-sector expressions in (\ref{Bbf})
share the same properties as their $A$-sector counterparts \cite{ADT,AD}.
First, they actually preserve the spacetime supersymmetry,
for $B^{\rm boson}$ and $B^{\rm fermion}$ are found to be equal
as functions of $\tau$.
Second, these expressions also exhibit the same minus-sign cancellations
that reduce the apparent central charge from $c=8$ to $c=6$.  Indeed,
in complete analogy to (\ref{prodA}), this central-charge
reduction exists because these $B$-sector expressions satisfy
the remarkable identity
\beq
        B^{\rm boson}~=~
        B^{\rm fermion}~=~
       4\,q^{1/2}\, \prod_{n=1}^\infty
     { (1+q^{3n})^4 \,(1-q^{3n})^2\over (1-q^n)^6 } ~.
\label{prodB}
\eeq
Note that the right side of this equation
is simply the $q$-expansion of $\thetatwo^2(3\tau)/\eta^6(\tau)$.
Therefore, just as the contributions from the $A$-sector
states were isomorphic to those of
free worldsheet bosons and fermions (where such fermions
are equivalent to internal bosons compactified on circles of radius $R=1$),
we see that the contributions from the $B$-sector states are also
isomorphic to free worldsheet bosons and fermions, but
with the fermions on rescaled compactification lattices
(or equivalently, with
the internal bosons compactified on circles of radius $R=\sqrt{3}$
rather than $R=1$) \cite{AD,D}.

Thus, the appearance of these new $B$-sector states
suggests the existence of additional
purely massive BPS states, fully consistent with $c=6$, but
beyond those described in Ref.~\cite{DVV}.

\setcounter{footnote}{0}
\section{Massless States and the Internal Projection}

The next step is to interpret the partition function
(\ref{partfunct}) that we have constructed, and in particular to
construct a worldsheet theory for which this function emerges as the
partition function.  After all, while this partition function has been
constructed with a number of properties in mind and on the basis of
several remarkable identities, it is {\it a priori}\/ far from clear
that there exists a consistent worldsheet theory underlying it.
In this section, we will give evidence
that these partition functions are consistent with a theory
whose massless states are those of the six-dimensional Green-Schwarz
string (as required), and also give evidence that
the internal projection is ultimately consistent.

Although we have constructed the partition function (\ref{partfunct})
in order to satisfy certain properties expected of the six-dimensional
self-dual string,
it remarkably turns out that this partition function
is precisely the partition function that was derived several
years ago for the so-called $K=4$ fractional superstring.
Detailed discussions concerning these fractional
superstrings can be found in the original
references \cite{AT,DT,ADT,ALT,FL,CR,AD,D}.
Note that the $K=4$ fractional superstring was originally
constructed as a generalization of the ordinary superstring or
heterotic string,
but with critical spacetime dimension $D_c=6$.
As such, in its original formulation,
it was a theory of gravity and contained a graviton.
However, as we shall see, it is possible to modify the original
interpretation of these partition functions in such a way as to yield
a suitable worldsheet formulation of the self-dual string.
Hence, although they share the same partition functions, these
new candidate strings are not to be identified with the original
fractional superstrings, and knowledge of fractional superstrings
will not be needed for what follows.

\subsection{The massless sector}

We shall concentrate on the left-moving worldsheet modes, since
the left- and right-moving modes can ultimately be combined to
form the states of the closed string.
Before the internal projection, we have seen that we can
describe the Fock space of our six-dimensional string theory in terms
of its pre-projection $c=8$ conformal field theory (CFT), namely a
tensor product of four free bosons along with four copies of the $\IZ_4$
parafermion theory:
\beq
   {\rm original~CFT} ~=~
     \left\lbrace \mathop\otimes_{i=1}^{4}  X^i \right\rbrace
          ~~\otimes~~
     \left\lbrace \mathop\otimes_{a=1}^{4}  (\IZ_4~{\rm PF})^a
     \right\rbrace~.
\label{originalCFT}
\eeq
This means that every state with spacetime mass $M$
in the Fock space can be written in the form
$\phi_h |0\rangle$  where $|0\rangle$ is the vacuum state
of the original CFT (\ref{originalCFT}),
and where $\phi_h$ is a field in this theory of conformal dimension $h$.
Note that $M^2=h-c/24= h-1/3$.

As indicated in (\ref{originalCFT}),
the four transverse coordinate bosons $X^i$  carry a spacetime
vector index $i$ spanning the  $(\bf 2,2)$ representation
of the $SU(2)\times SU(2)$ massless little group.
The zero modes of these bosons are then interpreted as the
transverse momenta of the states.
By contrast, the fields of each parafermion theory are assigned a
spacetime spinor index $a$ transforming in the  $2(\bf 2,1)$ representation
of $SU(2)\times SU(2)$.

With a little knowledge of the $\IZ_4$ parafermion conformal field theory,
it is easy to construct the lightest states which contribute to the
partition function (\ref{partfunct}).  As mentioned above,
the $\IZ_4$ theory
is particularly simple to work with because it has $c=1$, and is
equivalent to a free boson $\phi$ compactified on a circle of radius
$R=\sqrt{3/2}$.
The primary fields and conformal dimensions corresponding to
the $\IZ_4$ characters $\chi^{0}_{0}$, $\chi^2_2$,
$\chi^2_0$, and $\chi^4_2$ are thus
respectively identified as
\beqn
    {\bf 1}~&\Longleftrightarrow&~  h=0 ~,\nonumber\\
    \exp\left(\pm i\phi/\sqrt6\right) ~\equiv~\sigma_\pm
{}~&\Longleftrightarrow&~
	 h=1/12 ~,\nonumber\\
    \exp\left(\pm 2i\phi/\sqrt6\right)  ~\equiv~\epsilon_\pm
	~&\Longleftrightarrow&~h=1/3  ~,\nonumber\\
    \exp \left(\pm 3i\phi/\sqrt6\right)  ~\equiv~\psi_\pm
	~&\Longleftrightarrow&~h=3/4~.
\label{Zfour}
\eeqn
Note that the $\sigma_\pm$, $\psi_\pm$, and $\epsilon_\pm$ fields
are respectively the spin operator, parafermion current, and energy operator
in the $\IZ_4$ parafermion theory.

Given this identification of fields, it
is straightforward to determine the massless states that contribute
to the partition function (\ref{partfunct}).
Note that
massless states can only contribute
to the term $4 (\chi^0_0)^3\chi^2_0$ within $A^{\rm fermion}$,
or the term $4 (\chi^2_2)^4$ within $A^{\rm boson}$.
The corresponding massless states in the pre-projection conformal field
theory (\ref{originalCFT}) from $A^{\rm fermion}$ are thus
identified as
\beq
	|f\rangle ~=~ (\epsilon_\pm)^a_{-1/3}\,|0\rangle ~
\eeq
where the $\epsilon_\pm$ fields are defined in (\ref{Zfour}) and their
modings are given as a subscript.
These unusual modings follow from the fractional spin of this parafermionic
field, and are just what is required in order to create a massless state.
As before, the superscript denotes the spacetime spinor index
assigned to each parafermion factor.
Thus, these states describe eight spacetime fermionic degrees of freedom
transforming as four Weyl fermions, and fill out the  $4(\bf 2,1)$
representation
of the little group $SU(2)\times SU(2)$.

In a similar fashion, the massless states contributing to $A^{\rm boson}$ are
identified as
\beq
	|b\rangle ~=~ (\sigma_\pm)^1_{-1/12} (\sigma_\pm)^2_{-1/12}
	(\sigma_\pm)^3_{-1/12} (\sigma_\pm)^4_{-1/12} \,|0\rangle .
\eeq
To determine the spacetime interpretation of these states, we observe that
when acting on $|b\rangle$, the parafermionic
$\tilde\epsilon\equiv \epsilon_+ + \epsilon_-$
fields remarkably have zero modes which satisfy a Clifford algebra \cite{ALT}:
\beq
	\{ \tilde \epsilon^a_0, \tilde \epsilon^{a'}_0 \} \,|b\rangle
           ~=~ \delta^{a,a'} \,|b\rangle~.
\eeq
Note that this is a Clifford algebra involving spacetime {\it spinor}\/
indices, as in the Green-Schwarz string.
This implies that the massless $|b\rangle$ states fill out a representation
of this algebra.  It is easy to see that the spacetime Lorentz properties of
such
a representation are $8({\bf 1,1}) \oplus 4(\bf 2,1)$, thus describing
eight scalar bosons and four Weyl fermions.
We suppose that these fermionic states are removed by an analogue of the
GSO projection in the Ramond sector of the Neveu-Schwarz-Ramond superstring.
This is consistent with the spacetime supersymmetry of the partition
function.

Thus, we have identified the massless states in the pre-projection
conformal field theory, and find that they fall into massless hypermultiplets
of
six-dimensional $N=(1,0)$ supersymmetry.  Upon combining left- and
right-moving sectors in a chiral way, we obtain massless states in
the tensor multiplet of $N=(2,0)$ supersymmetry.
Note that this description of the self-dual string is
a peculiar mixture of the Green-Schwarz and Neveu-Schwarz-Ramond
formulations of the ten-dimensional superstring:
on the one hand some worldsheet fields carry spacetime
spinor quantum numbers, and on the other hand a GSO-like
projection removes the tachyonic as well as some of the massless
states from the spectrum.

\subsection{The internal projection}

We now discuss the effects of the ``internal projection''
which removes states from the spectrum of this theory.
Recall that this internal projection is associated with the minus
signs appearing within the expressions $A^{\rm boson,fermion}$ and
$B^{\rm boson,fermion}$ in (\ref{Abf}) and (\ref{Bbf}).
In particular, note that these minus signs
precede the characters of sectors which contain only massive
states.  Thus, this internal projection
only affects the {\it massive}\/ string states, which were the source
of Lorentz symmetry-breaking in the usual formulation of the six-dimensional
Green-Schwarz string.
Moreover, since this internal projection removes only
massive states, the description of the massless states given above
remains unaffected.

Although this projection may seem similar to the ordinary
GSO projection,
it has some unusual features.
For example, if we arrange the states of the
ordinary superstring into towers of conformal descendents of primary
states (Verma modules), then the GSO projection in the ordinary
superstring either leaves a given tower intact, or or projects it out
entirely.  By contrast, in the self-dual string partition
function (\ref{partfunct}), the new internal projection which
appears projects out only {\it some}\/ of the
states in each individual tower, leaving behind a seemingly random set
of states which therefore cannot be interpreted as the complete Fock
space of the original underlying worldsheet CFT.

Ordinarily, this would seem to render the
spacetime spectrum of this theory inconsistent
with any underlying worldsheet-theory interpretation.
Remarkably, however, evidence suggests \cite{AD,D} that the residual states
which
survive the internal projection in each tower precisely recombine to
fill out the complete Fock space of a {\it different}\/ underlying
CFT. Thus, whereas the ordinary GSO projection merely removes
certain highest-weight sectors of the worldsheet conformal field theory,
this new internal projection appears to actually change the
underlying conformal field theory itself.
In fact, as we have seen above, the effective
central charge of the resulting post-projection
CFT is smaller than that of the
original (parafermionic) CFT.
Thus, the internal projection
removes exponentially large numbers of states from each
mass level of the original Fock space (so as to alter the asymptotic
behavior of the degeneracy of states).
Indeed, such a drastic projection
somewhat resembles the BRST projection
which enables unitary minimal models with $c<1$ to be constructed
from free $c=1$ bosons in the Feigin-Fuchs construction.

In practical terms, the appearance of this internal projection means
the following.
Although the original worldsheet theory was described
by (\ref{originalCFT}), we see that
the internal projection deforms this theory
into a smaller theory with $c=6$:
\beq
  {\rm new~CFT}~=~
     \left\lbrace \mathop\otimes_{i=1}^{4}  X^i \right\rbrace
          ~~\otimes~~
   \left\lbrace {\rm internal}~c{=}2~{\rm theory}\right\rbrace~.
\label{finalCFT}
\eeq
This not only means that some of the original states $\phi_h|0\rangle$
are projected out of the spectrum, but also that those states which
survive can equivalently be written in the form
$\phi'_{h'}|0\rangle_{\rm new}$ where
$|0\rangle_{\rm new}$ is the new vacuum state of the post-projection
conformal field theory (\ref{finalCFT}),
where $\phi'_{h'}$ is a field of conformal dimension $h'$
in this new CFT,
and where the spacetime mass $M$ can
be equivalently identified as $M^2= h'-c'/24=h'-1/4$.

Verifying that the internal projection leaves behind a self-consistent
CFT is a difficult task, since it requires detailed knowledge of
precisely which original parafermionic states $\phi_h|0\rangle$ are
projected into or out of the spectrum.  This in turn requires an
internal-projection operator, constructed out of the
parafermionic fields, which would enable us to analyze these
projections at the level of individual states.
Unfortunately, despite
certain clues involving the so-called parafermionic ``twist current''
\cite{CR,AD,D}, no such internal-projection operator has been
constructed.

In some sense, then, this worldsheet theory
is in a position that can best
be appreciated by imagining that the history of the ordinary superstring
had been different,
and that modular invariance had been discovered {\it before}\/ the
discovery of the GSO projection.  Starting with the RNS model,
one would have calculated the full (unprojected) RNS spectrum, and then
would have constructed a modular-invariant partition function.
One would have observed, in this partition function, a minus sign
that signalled a projection on the Fock space of states.  This would have
seemed mysterious, and it would only later have been discovered
that the correct way to implement this projection would be through
the $G$-parity operator.
In the present case, we have calculated
a unique partition function, and have observed some unexpected minus signs.
In our case, however, these signs indicate not an ordinary GSO projection,
but rather a mysterious internal projection.

The most compelling evidence to date for the
consistency of the internal projection been obtained through a detailed
analysis of the partition function.  This
evidence is examined in detail in Ref.~\cite{D}, where a full
discussion of the known properties
(central charge, highest weights, fusion rules, and characters)
of the post-projection CFT is
given.  Non-technical discussions can also be found in
Refs.~\cite{Reviewone,Reviewtwo}.
However, actually constructing a suitable representation for
the post-projection conformal field theory in terms of worldsheet
fields remains an open question.

\subsection{Self-dual strings in lower dimensions}

Along with the six-dimensional non-critical self-dual string,
it has been pointed out \cite{Wit3}
that there also exist
four-dimensional $N=2$ supersymmetric
string vacua
whose low-energy excitations include light (and tensionless) strings.
Therefore, just as with the
six-dimensional self-dual string, it may be conjectured that
there exists a worldsheet formulation of this lower-dimensional
string as well.

One of the advantages of the parafermionic formulation we have proposed
above is that it readily generalizes to four and three spacetime
dimensions \cite{AT,ADT}.
In particular, the four-dimensional analogue of the above partition
function consists of the characters of two transverse coordinate
bosons along with those of two $\IZ_8$ parafermions, while the
three-dimensional analogue consists of the characters of a single
transverse coordinate boson along with that of a single
$\IZ_{16}$ parafermion.
In each case, the resulting partition functions are consistent
with spacetime supersymmetry, and have the appropriate central
charges $c=3$  and $c=3/2$ respectively.
Just as in the six-dimensional case,
these lower-dimensional central charges are obtained
in an analogous fashion as the result of an internal projection on
the corresponding parafermionic Fock spaces.
These lower-dimensional partition functions also contain
additional unexpected sectors which are completely analogous
to the $B$-sectors in the six-dimensional case, and which
satisfy identities analogous to (\ref{prodB}).
Details can be found in Refs.~\cite{AD,D}.

Furthermore,
note that there also exist $N=(1,0)$ versions of the six-dimensional
self-dual string theories \cite{DMW,GH,SeiWit,Ganor2}.
These may very well be realized as heterotic-like
constructions involving the parafermionic worldsheet theories
we have proposed here.
Indeed, modular-invariant partition functions of this nature
have already been constructed \cite{DT}, and may be interpreted
in a similar manner.

\section{Discussion}

In this note, we have proposed a framework for a worldsheet theory which
might be a reasonable
candidate for the worldsheet theory of the six-dimensional self-dual string.
As far as we are aware,
our proposal is presently the only known
six-dimensional candidate theory which
is spacetime supersymmetric, has central charge $c=6$,
and gives rise to a partition function consistent with
(\ref{expectedpartfunct}).

Due to their unusual parafermionic
formulation, however, we see that
the method by which quantum consistency is restored
in these worldsheet theories
is quite different than the solution proposed in Ref.~\cite{Schwarz}
for the Green-Schwarz string:
rather than introduce extra excitation modes (which increase the central
charge beyond its desired value), these parafermionic worldsheet theories
instead give rise to new sectors with highly unusual properties and
to an internal projection which deforms the underlying parafermionic Fock
space.  Such mysterious sectors and projections may therefore prove to
be the key to understanding the worldsheet physics of the conjectured
six-dimensional self-dual string.

Of course, we do not understand this theory very well yet.  Indeed, as we
have discussed above, we only can only represent this theory
by starting from a (larger) parafermionic CFT formulation, followed by a
highly non-trivial internal projection.  Is there a direct formulation of this
theory?  This remains the major question.  Thus, given the present
state of knowledge concerning either the self-dual string or the
post-projection light-cone gauge worldsheet CFT we are constructing,
it is not yet possible to make a more concrete test of this proposal.
Such issues await a proper formulation and understanding
of the internal projection, and a proper elucidation of the role played
by the extra parafermionic $B$-sectors.

\setcounter{footnote}{0}
\bigskip
\bigskip
\leftline{\large\bf Acknowledgments}
\bigskip

We are pleased to thank
T. Banks, J. Blum, O. Ganor, A. Hanany, R. Myers, A. Peet,
J. Schwarz, N. Seiberg, S. Shenker, and E. Witten
for discussions.
This work was supported in part by
DOE grants
DE-FG05-90ER40559
and
DE-FG05-90ER40542.

\vfill\eject
\bibliographystyle{unsrt}

\end{document}